\documentclass[prl,amsmath,amssymb,twocolumn,showpacs,floatfix,groupedaddress,superscriptaddress]{revtex4}
\usepackage{graphicx}
\usepackage[english]{babel}
\usepackage{times}
\usepackage{dcolumn}
\usepackage[usenames,dvipsnames]{color}
\usepackage{units}

\begin{document}

\title{Lifshitz phase transitions in the ferromagnetic
regime of the Kondo lattice model}

\author{Denis Gole\v{z}}

\affiliation{Jo\v{z}ef Stefan Institute, Jamova 39, SI-1000 Ljubljana,
Slovenia}
\author{Rok \v{Z}itko}

\affiliation{Jo\v{z}ef Stefan Institute, Jamova 39, SI-1000 Ljubljana,
Slovenia}
\affiliation{Faculty for Mathematics and Physics, University of
Ljubljana, Jadranska 19, SI-1000 Ljubljana, Slovenia}

\date{\today}

\begin{abstract}
We establish the low-temperature phase diagrams of the spin-$1/2$ and
spin-$1$ Kondo lattice models as a function of the conduction-band
filling $n$ and the exchange coupling strength $J$ in the regime of
ferromagnetic effective exchange interactions ($n \lesssim 0.5$). We
show that both models have several distinct ferromagnetic phases
separated by continuous Lifshitz transitions of the Fermi-pocket
vanishing or emergence type: one of the phases has a true gap in the
minority band (half metal), the others only a pseudogap. They can be
experimentally distinguished by their magnetization curves; only the
gapped phase exhibits magnetization rigidity.
We find that, quite generically, ferromagnetism and Kondo screening
coexist rather than compete, both in spin-$1/2$ and spin-$1$ models.
We compute the Curie temperatures and establish a ``ferromagnetic
Doniach diagram'' for both models.
\end{abstract}


\maketitle

\newcommand{\Schroedinger}{Schr\"{o}dinger }
\newcommand{\ave}[1]{\langle #1 \rangle}
\newcommand{\bra}[1]{\langle #1|}
\newcommand{\green}[2]{\langle \langle #1 , #2 \rangle\rangle}

\newcommand{\vc}[1]{{\mathbf{#1}}}
\newcommand{\braket}[2]{\langle#1|#2\rangle}
\newcommand{\expv}[1]{\langle #1 \rangle}
\newcommand{\corr}[1]{\langle\langle #1 \rangle\rangle}
\newcommand{\ket}[1]{| #1 \rangle}
\newcommand{\Tr}{\mathrm{Tr}}
\newcommand{\kor}[1]{\langle\langle #1 \rangle\rangle}
\newcommand{\degg}{^\circ}
\renewcommand{\Im}{\mathrm{Im}}
\renewcommand{\Re}{\mathrm{Re}}
\newcommand{\GG}{{\mathcal{G}}}
\newcommand{\atanh}{\mathrm{atanh}}
\newcommand{\sgn}{\mathrm{sgn}}
\newcommand{\beq}[1]{\begin{equation}#1\end{equation}}

Materials with competing interactions, such as many lanthanide and
actinide compounds, have complex low-temperature phase diagrams with
different ground states
\cite{Coleman:2005gb,Lohneysen:2007ve,Si:2010ef,
Stewart:2001zz,Gegenwart:2008vc,Pfleiderer:2009gc}. 
The Kondo lattice model (KLM)
\cite{Tsunetsugu:1997vw,Gulacsi:2004kk,hewson} describes a
conduction band of itinerant electrons and a lattice
of local moments on $f$ shells, coupled at
each site by an antiferromagnetic exchange interaction $J$. For large
$J$, the moments are screened. The resulting paramagnetic state
has Fermi liquid properties with strongly renormalized parameters.
For small $J$, the conduction-band electrons are
carriers of long-range magnetic interactions
and the moments order. The two regimes are separated by a
quantum phase transition at critical $J^*$, as
described by the Doniach diagram \cite{doniach1977}.
The N\'eel temperature increases at first quadratically with $J$, 
but then it peaks and decreases to zero at $J^*$ as the Kondo
screening takes over. 
The simplest version of the KLM with spin-$1/2$ moments indeed
has an antiferromagnetic (AFM) ground state (N\'eel order) for small
$J$ near half-filling
\cite{Capponi:2001jc,otsuki2009}. The nature of the phase
transition at $J^*$ has been investigated using a variety
of methods, the most accurate of which confirm that the transition is
second order (quantum critical) and indicate that it involves a change
of the Fermi surface topology
\cite{DeLeo:2008jy,Martin:2008in,Martin:2010iv}. In the spin-$1$ KLM,
there is no phase transition at half-filling and the AFM phase extends
to large values of $J$.

While most cerium compounds show AFM order, some are ferromagnetic
(FM): CeRu$_2$Ge$_2$ \cite{Sullow:1999vh}, CeIn$_2$
\cite{Rojas:2009bx,Mukherjee:2012ju}, and CeRu$_2$Al$_2$B
\cite{Baumbach:2012bd}. A number of uranium and neptunium
heavy-fermion materials are also FM: UTe \cite{Schoenes:1984vy},
UCu$_{0.9}$Sb$_2$ \cite{Bukowski:2005fb}, UCo$_{0.5}$Sb$_2$
\cite{Tran:2005jw}, NpNiSi$_2$ \cite{Colineau:2008hx}, Np$_2$PdGa$_3$
\cite{Tran:2010kn}, and UCu$_2$Si$_2$ \cite{Troc:2012kt}. In addition,
there are strong indications of robust coexistence of the Kondo effect
and ferromagnetism, in particular in U compounds. In
Refs.~\cite{Perkins:2007ez,Perkins:2007jt,Coqblin:2009fl,Thomas:2011jn,Troc:2012kt}
it has been proposed that an appropriate minimal model for this
behavior is the spin-$1$ version of the KLM, where in the mean-field
picture the conduction-band electrons underscreen the local moments,
while the residual moments order ferromagnetically. FM order appears
for low and moderate electron filling $n$ in the conduction band, $n
\lesssim 0.5$
\cite{Lacroix:1979uo,Batista:2002ip,Perkins:2007ez,peters2007magnetic,otsuki2009b}.
Mean-field analysis predicts two phases: for small $J$ the stable
phase is a FM regular metal, while for large $J$ there is a transition
to a FM heavy metal. Dynamical mean-field theory (DMFT) calculations
demonstrated that the spin-$1/2$ KLM also has a FM order coexisting
with (incomplete) Kondo screening \cite{peters2012}. Furthermore, this
phase is a half-metal with gapped minority-spin band and a
commensurability condition relates the magnetization to filling $n$
\cite{peters2012} due to completely filled minority-spin lower band
\cite{beach2008,kusminskiy2008}. A recent mean-field analysis of the
spin-$1/2$ model suggested the presence of several different
ferromagnetic phases \cite{Liu:2013uj}. So far, however, a single FM
phase has been identified in the DMFT calculations
\cite{peters2007magnetic,otsuki2009b}. 

These findings open a number of questions: What is the
relationship between ferromagnetism and Kondo screening: do they
compete or coexist? 
What is the minimal model for studying these effects, spin-$1/2$ or
spin-$1$ KLM? Is there a quantum phase transition between different FM
states also in the spin-$1/2$ model? What is the nature of these
transitions and what are their experimental signatures? And, finally,
which aspects of the static mean-field analysis \cite{mf} are correct
and which must be revised in more accurate dynamical treatment? To
answer these questions we have performed extensive DMFT
\cite{georges1996} calculations using the numerical renormalization
group (NRG) as the impurity solver
\cite{wilson1975,bulla2008,hofstetter2000,
peters2006,weichselbaum2007,resolution}, as well as static mean-field
calculations for both models \cite{mf}.

We consider the Kondo lattice model 
\begin{equation}
\begin{split}
\mathcal{H} &= \sum_{\vc{k}\sigma} (\epsilon_k-\mu)
c^\dag_{\vc{k}\sigma} c_{\vc{k}\sigma}
+ J \sum_i \vc{s}_i \cdot \vc{S}_i, \\
\end{split}
\end{equation}
which describes a single-orbital conduction band with dispersion
$\omega=\epsilon_k$, and a lattice of local moments described by the
spin-$S$ operators $\vc{S}_i$; $\vc{s}_i$ is the conduction-band
spin-density at site $i$, and $J$ is the antiferromagnetic Kondo
exchange coupling ($J>0$). We focus on the Bethe lattice that has a
semicircular density of states with bandwidth $2D$.

\begin{figure}
\centering
\includegraphics[width=8cm]{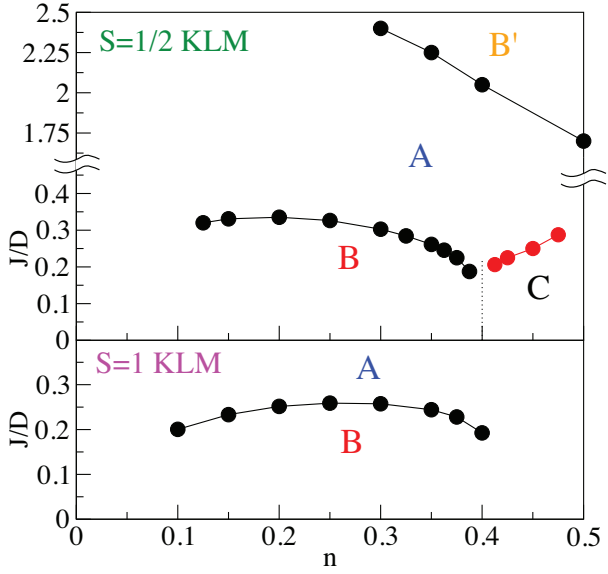}
\caption{(Color online) Phase diagrams of spin-$1/2$ and spin-$1$
Kondo lattice models for $n<0.5$. Phase A is a ferromagnetic
half-metal phase with strong Kondo effect where the minority band is
gapped. Phases B and B' are itinerant ferromagnetic phases with a pseudogap.
Phase C for spin-1/2 model indicates the region with charge order
\cite{peters2013}.
For very small $n$, the calculations fail to converge.
} \label{fig1}
\end{figure}

\begin{figure}
\centering
\includegraphics[width=8cm]{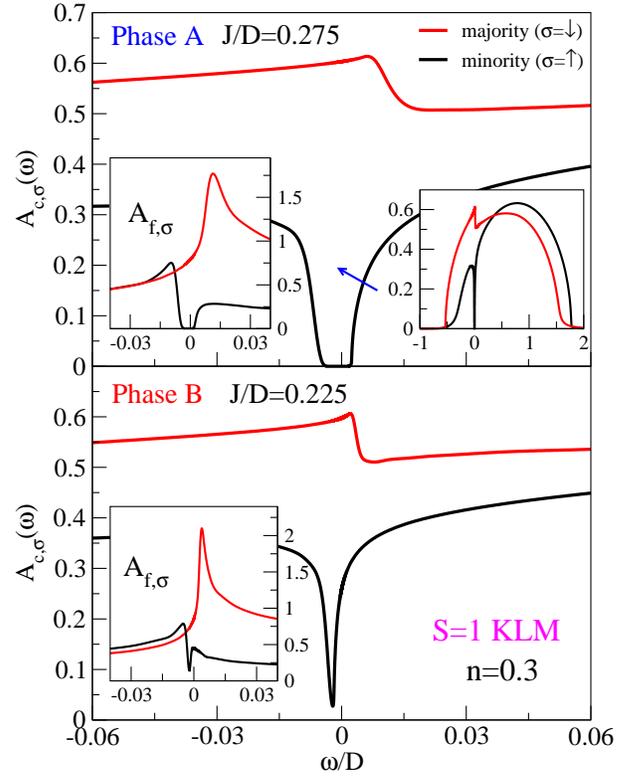}
\caption{(Color online) Spin-resolved conduction-band local spectral
functions $A_{c,\sigma}$ for the spin-$1$ KLM in the ferromagnetic
half-metal phase (A) and in the itinerant ferromagnetic phase (B). The
arrow indicates the main effect of decreasing interaction $J$: the
lower edge of the upper hybridized band shifts to lower frequencies.
The left insets in both panels show the $f$-level spectral functions
$A_{f,\sigma}$ defined through the imaginary part of the scattering
$T$ matrix. The right inset in the upper panel shows the spectral
functions in the full frequency interval.} \label{fig2}
\end{figure}

In Fig.~\ref{fig1} we present the main result of this work: the phase
diagrams of the spin-$1/2$ and spin-$1$ KLM as a function of $n$ and
$J$. For {\it both} spins we find several different ferromagnetic
phases. Phase A corresponds to the ferromagnetic half-metal phase
described by Peters et al. \cite{peters2012}. The corresponding
spin-resolved spectral functions for the $S=1$ model are shown in
Fig.~\ref{fig2}, panel A. The minority spin band is gapped
\cite{peters2012}, while the majority band exhibits the weak
hybridization pseudo-gap characteristic of the Kondo lattice systems
\cite{pruschke2000,costi2002}. Phase B at small $J$ is not gapped, but
there is a pronounced pseudogap just below the Fermi level in the
minority band, Fig.~\ref{fig2}, panel B. The spectral functions for
the $S=1/2$ model are qualitatively the same.
The spectra thus suggest the occurrence of a Lifshitz transition at
$J^*$: there is no change in the symmetry, but the Fermi surface of
the minority band shrinks to a point and disappears as one goes from
phase $B$ to $A$. We emphasize that the two phases exist both for
spin-$1/2$ and for spin-$1$ models and have similar properties;
clearly, within the DMFT, the {\it value of the spin does not play a
crucial role in the BA transition}.  $J^*$ is a non-monotonic function
of $n$ that peaks at $n \sim 0.2$ and $n \sim 0.25$, respectively.
Near $n \sim 0.4$ we observe change of behavior in the small-$J$
phase. For $S=1/2$ KLM, this is the parameter regime where charge
order occurs \cite{otsuki2009b,peters2013}, but it is not allowed for in our
calculations.

\begin{figure}
\centering
\includegraphics[width=8cm]{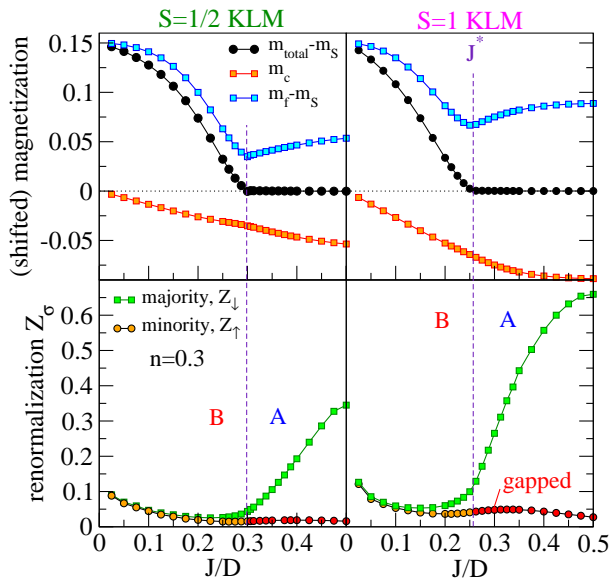}
\caption{(Color online) Total, conduction-band $c$-level and localized
$f$-level magnetizations (top panels) and the spin-dependent
quasiparticle renormalization factors $Z_\sigma$ (bottom panels)
across the phase transition, indicated by the vertical dashed lines.
The magnetization is here defined as the expectation value of the
spin operator:
$m_f=\expv{S_z}$, $m_c=(n_\uparrow-n_\downarrow)/2$,
$m_\mathrm{total}=m_f+m_c$. In the plots, $m_\mathrm{total}$ and
$m_f$ are shifted by $m_S$ defined in Eq.~\eqref{mS}.
} \label{fig3}
\end{figure}

In Fig.~\ref{fig3} we plot the magnetization and the quasiparticle
renormalization factor $Z_\sigma=[1-\mathrm{d}
\Sigma_\sigma/\mathrm{d}\omega(\omega=\mu)]^{-1}$
as a function of $J$ across the BA transition. The frozen
magnetization in phase A is given by a generalization of the
spin-$1/2$ KLM result from
Refs.~\cite{beach2008,kusminskiy2008,peters2012}:
\begin{equation}
\label{mS}
m_S = (2S-n)/2.
\end{equation}
At transition, the magnetization is continuous with a change of
slope in $m_f$. This is in disagreement with the static mean-field
analysis for $S=1$ which predicts a jump 
\cite{Perkins:2007jt}. The factors $Z_\sigma$ for both spin
orientations are continuous and finite across the transition (in the
minority band of phase A there are no quasiparticles, but $Z_\sigma$
can formally still be defined). There is thus no criticality in this
spin-selective metal-insulator transition, which may be identified as a {\it continuous
Lifshitz transition of the Fermi pocket vanishing type}
\cite{lifshitz1960,Yamaji:2006kc,kusminskiy2008,beach2008,Li:2010ih,Bercx:2012vx}.
The Fermi surface topology is continuous with no reorganization. Deep
in the phase A, the majority electrons become weakly correlated ($Z$
has a value of order $0.5$).

For very large $J$, in the spin-$1/2$ model (but not for spin-$1$)
there is another Lifshitz transition to a non-gapped phase
\cite{robert} that we denote as B'. While in the BA transition, the
chemical potential is located at the {\it bottom of the upper
hybridized band}, in the AB' transition the chemical potential is
located at the {\it top of the lower hybridized band} at the
transition point. In other words, while BA corresponds to the
vanishing of electron pocket, AB' corresponds to the emergence of hole
pocket. 
For even larger $J$, the system eventually becomes paramagnetic (for
$n=0.3$ at $J/D=3.4$).

The static mean-field theory for $S=1/2$ also predicts distinct phases
\cite{Liu:2013uj,mf} which roughly correspond to B, A, and B'. The
exact treatment of quantum fluctuations in DMFT leads, however, to a
number of differences: i) The small-$J$ phase B is not pure
ferromagnetic, but there is a coexistence with the Kondo effect. In
the static MF treatment only pure ferromagnetic solution is stable and
the phase transition from the corresponding phases A to B is of the
first order \cite{Li:2010ih}, for details see Supplementary materials
\cite{mf}. Small-$J$ phase B is not pure ferromagnetic. ii) The
Lifshitz transitions are all continuous: there are no jumps in any of
the results. iii) Deep inside phases B and B' there are pseudo-gaps
rather than gaps. This is due to non-zero imaginary part of the
self-energy in DMFT, i.e., due to correlation effects. The most
surprising outcome of the DMFT calculations is, in fact, the gradual
emergence of true gaps from pseudo-gaps as the gapped phase A is
approached from B or from B', while the static MF results are closer
to the rigid-band picture.

\begin{figure}
\centering
\includegraphics[width=8cm]{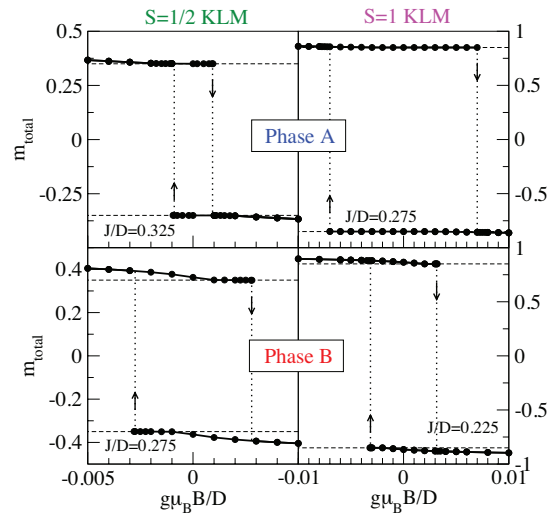}
\caption{(Color online) Hysteresis loops: magnetization in
longitudinal external magnetic field. The dashed lines indicate the
value of the frozen magnetization $m_S$. The $g$-factors are assumed
equal for $c$ and $f$ levels, $g_c=g_f=g$. Occupancy is $n=0.3$.} \label{figM}
\end{figure}

Does the existence of multiple phases indicate a competition between
the exchange interaction and the Kondo effect? Some degree of antagonism
is suggested by the fact that the $f$-shell magnetization $m_f$ has a
minimum at the BA Lifshitz point where both tendencies are expected to
be equally strong and, furthermore, it could be argued that $m_f$
increases with $J$ in phase A only because Kondo screening is rendered
incomplete by the opening and widening of the gap. Nevertheless, this
competition does not imply mutual exclusion and most results rather
support the notion of robust coexistence.

Experimentally the phases can be distinguished by their
magnetization curves. In phase A, $m_\mathrm{total}$ remains pinned to
$m_S$ for a finite range of the field strength, while in phase B the
susceptibility $dM/dB$ near zero field is finite, see Fig.~\ref{figM}.
For sufficiently strong field, a gap opens in the minority band in
phase B, too. This effect can be understood within a rigid-band
picture.
For very strong field, the magnetization is reoriented in a
first-order spin-flop transition which preempts another Lifshitz
transition.

\begin{figure}
\centering
\includegraphics[width=8cm]{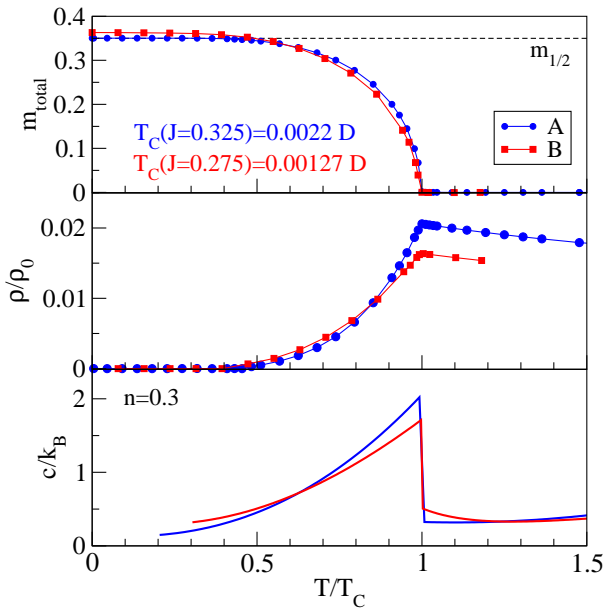}
\caption{(Color online) Temperature dependence of the magnetization,
resistivity and heat capacity for the spin-$1/2$ Kondo lattice model
in phases A and B. The horizontal axis is rescaled by the Curie
temperature $T_C$. Resistivity is in units of $\rho_0=2\pi e^2
\Phi(0)/\hbar D$, where $\Phi$ is the transport integral. Heat
capacity curve was obtained by differentiating a piecewise
interpolation of the numerical results for the total energy. }
\label{fig4}
\end{figure}

In Fig.~\ref{fig4} we plot the temperature dependence of key
thermodynamic and transport properties in phases A and B. 
We find that the magnetization in phase B remains essentially pinned
at $m_S$ until $T$ becomes of the order of the gap, while it has a
finite temperature-derivative at $T=0$ in phase A. This difference is,
however, small. The resistance $\rho$ increases in both phases up to
the Curie temperature $T_c$, then it decreases approximately as a
power-law $T^{-0.3}$, not logarithmically. The heat capacity $c$ has a
jump discontinuity at $T_c$. Similar features are indeed observed
experimentally, for example in
Refs.~\cite{Tran:2005jw,Colineau:2008hx,Baumbach:2012bd}, although the
simple KLM does not capture the full complexity of real materials.

\begin{figure}
\centering
\includegraphics[width=8cm,clip]{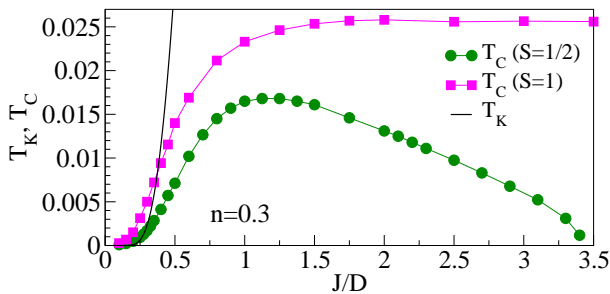}
\caption{(Color online) ``Ferromagnetic Doniach diagram'' for
spin-$1/2$ and spin-$1$ Kondo lattice models. 
} \label{fig5}
\end{figure}

We summarize the behavior of both Kondo lattice models in the form of
a ``ferromagnetic Doniach diagram'' in Fig.~\ref{fig5}. We plot the
Kondo temperature for a single-impurity model with flat band (which
does not depend on the impurity spin \cite{andrei1983}) and the Curie
temperature $T_C$ for each model. The Curie temperature has no
observable feature at the Lifshitz transition points $J^*$. Apart from
the (approximately) factor of two difference, there is no difference
in $T_C$ of spin-1/2 and spin-1 models for small $J$. At large $J$,
spin-1/2 model first goes into the B' phase and then becomes
paramagnetic. The spin-1 model remains ferromagnetic in the large $J$
limit. This is similar to the behavior of the AFM phases of both
models at half filling.

We conclude by answering the questions raised in the introduction. 
There is no Kondo breakdown and no criticality, but rather a
continuous filling of the lower minority band and the disappearance of
the electron pockets (and the emergence of hole pockets in the
spin-1/2 model for large $J$). We find robust coexistence of FM order
and Kondo screening in all phases, for both spins.
Kondo underscreening does not need to be invoked to explain the
magnetic ordering. Both models have qualitatively the same phase
diagram for physically most relevant small $J$. The Lifshitz
transitions are observable in the temperature and magnetic-field
dependence of the magnetization.
The static mean-field appears to be valid at the qualitative level,
however to properly describe the real nature of ferromagnetic phases
and transitions it is necessary to take into account dynamic effects,
as in the DMFT treatment.

\begin{acknowledgments}
We acknowledge discussions with Robert Peters and Janez Bon\v{c}a and
the support of the Slovenian Research Agency (ARRS) under Program
P1-0044.
\end{acknowledgments}


\clearpage
\newpage

\onecolumngrid
\appendix*

\begin{center} 
{\bf \Large Supplemental Material}
\end{center}

\section{Static mean-field theory}

\subsection{The $S=1/2$ case}

We perform a mean-field decomposition in the KLM written in the form:
\begin{equation}
\mathcal H = \sum_{k,\sigma} \epsilon_k c_{k,\sigma}^{\dagger} c_{k,\sigma} +
J \sum_i \vc{s}_i \cdot \vc{S}_i + \mu_B H  \sum_{i}  (g_c s_{z,i}+g_f
S_{z,i}),
\end{equation}
where $H$ is the external magnetic field oriented along the $z$ axis,
$\mu_B$ the Bohr magneton, while $g_{c}$ and $g_{f}$ are the Land\'{e}
factors. For simplicity, we consider flat non-interacting
conduction-band density of states (DOS): 
\beq{
\rho_c^0=1/2D,
}
where $D$ is the half-bandwidth.

The interaction term for localized spins with $S=1/2$ is
decomposed in terms of the hybridization operators \cite{beach2005,kusminskiy2008}
\beq{
\chi^{\mu}=\frac{1}{\sqrt{2}} \sum_{\alpha,\beta} f^{\dagger}_\alpha
\sigma^{\mu}_{\alpha\beta} c_{\beta},
}
where $c,f$ are annihilation operators for itinerant and localized
electrons, respectively, and the spin indexes $\alpha$ and $\beta$
range over spin up and down. The index $\mu$ ranges over $0,1,2,3$;
the operator $\sigma^{0}$ is the identity, while other $\sigma^i$ are
the Pauli matrices. These operators are complete in the spin sector
$1/2\otimes 1/2=1 \oplus 0$, and therefore the interaction part can be
split into: 
\begin{equation}
\vc{s} \cdot \vc{S} = 
\left( \frac{1}{2} c^{\dagger} \boldsymbol{\sigma} c \right)
\cdot
\left( \frac{1}{2} 
f^{\dagger} \boldsymbol{\sigma} f \right) = -3/4 \chi^{0\dagger}
\chi^{0} + 1/4 \boldsymbol{\chi}^{\dagger} \cdot \boldsymbol{\chi}.
\end{equation}
This expression is exact. 

We perform the standard mean-field procedure: $A
B\approx\ave{A}B+A\ave{B}-\ave{A}\ave{B}$. We assume that only the
singlet part $\ave{\chi^{0}}$ is nonzero and we use the $U(1)$ gauge
freedom to make $\ave{\chi_0}$ real. 

The second mean-field decomposition is done in the magnetic channel
(assuming the magnetization is along the $z$ axis):
\beq{
\vc{s} \cdot \vc{S}=s_z \tilde m_{f}+\tilde m_{c} S_z-\tilde m_{c} \tilde
m_{f},
}
where 
\beq{
\tilde m_{c}=\ave{s^z} \quad \text{and}\quad \tilde m_{f}=\ave{S^z}
}
are the expectation values of the $z$ component of conduction-band and
localized-electron spin. These are proportional to the magnetization
of $c(f)$ electrons:
\beq{
m_{f(c)}=-\mu_{B} g_{f(c)} \tilde m_{f(c)}.
}

In order to fix the average number of electrons we introduce the
chemical potential $\mu$. 
%
We also introduce Lagrangian multipliers $\lambda_i$ to enforce the
local constraint $\ave{n_{f,i}}=1$ on the $f$ electrons:
\beq{
\sum_i \lambda_i \sum_\sigma \left( f^\dag_{i,\sigma} f_{i,\sigma} -1
\right).
}
This constraint is fulfilled only as an average over all $f$
electrons, $\lambda_i \equiv \lambda$. We may then perform a FT:
\beq{
\lambda \sum_k \sum_\sigma \left( f^\dag_{k\sigma} f_{k\sigma} - 1
\right).
}
Thus $\lambda$ plays the role of the effective $f$ level energy: the
$f$ level occupancy is controlled by the difference between $\lambda$
and $\mu$.

At constant $\mu$, the thermodynamic potential that we need to
minimize is
\beq{
K(\mu,\ldots)=H(N_\mathrm{total},\ldots)-\mu N_\mathrm{total}=H-\mu (N_c+N_f) = H-\mu \sum_{k,\sigma} \left( c^\dag_{k\sigma} c_{k\sigma} + f^\dag_{k\sigma} f_{k\sigma} \right).
}

The mean-field thermodynamic potential takes the following wave-vector
representation:

\beq{
\mathcal K_{MF} = \sum_{k\sigma} \left( c^\dag_{k,\sigma}\,
f^\dag_{k,\sigma} \right) \tilde M_k \begin{pmatrix} c_{k,\sigma} \\ f_{k,\sigma}
\end{pmatrix} + \sum_k E_0,
}

where the  matrix $M_k$ is
%
%

\beq{
\tilde M_k = \begin{pmatrix} 
\epsilon_{k,\sigma}-\mu & -c \chi_0 \\
-c \chi_0 & \lambda_\sigma-\mu
\end{pmatrix}.
}

with
\beq{
\epsilon_{k,\sigma} = \epsilon_k + \epsilon_\sigma = \epsilon_k + J \tilde m_f \frac{\sigma}{2} +
\mu_B g_c H \frac{\sigma}{2} = \epsilon_k + \mu_B g_c \tilde H_c
\frac{\sigma}{2},
}
\beq{
\lambda_\sigma = \lambda + J \tilde m_c \frac{\sigma}{2} +
\mu_B g_f H \frac{\sigma}{2} = \lambda + \mu_B g_f \tilde H_f
\frac{\sigma}{2},
}
\beq{
c = \frac{3}{4} \frac{1}{\sqrt{2}} J = \frac{3J}{4\sqrt{2}},
}
\beq{
E_0 = +\frac{3}{4}J \chi_0^2-J \tilde m_c \tilde m_f-\lambda.
}
The effective field felt by the $c(f)$ electrons is given by 
\beq{
\tilde H_{c(f)}=H+\frac{J \tilde
m_{f(c)}}{\mu_B g_{c(f)}}.
}

In general, the equation of motion (EOM) can be written as
  \begin{equation}
  z \green{A}{B} = -\green{[\mathcal K_{MF},A]}{B}+\langle \langle [A,B] \rangle\rangle ,
  \end{equation}
where $A,B$ are arbitrary fermionic operators.
We find
\beq{
\begin{split}
z G_{cc,k\sigma} &= 1 + \left( \epsilon_{k\sigma}-\mu \right)
G_{cc,k\sigma} - c\chi_0 G_{fc,k\sigma}, \\
z G_{ff,k\sigma} &= 1 + \left( \lambda_\sigma-\mu \right)
G_{ff,k\sigma} -c\chi_0 G_{cf,k\sigma}, \\
z G_{cf,k\sigma} &= (\epsilon_{k\sigma}-\mu) G_{cf,k\sigma} - c\chi_0
G_{ff,k\sigma}, \\
z G_{fc,k\sigma} &= (\lambda_\sigma-\mu) G_{fc,k\sigma} - c\chi_0
G_{cc,k\sigma}.
\end{split}
}
Note also that $G_{cf}(z)=G_{fc}(z)$, since the matrix $\tilde M_k$ is
symmetric. It follows
\beq{
\begin{split}
(z-\epsilon_{k\sigma}+\mu)G_{cc,k\sigma} &= 1-c\chi_0 G_{fc,k\sigma}, \\
(z-\lambda_{\sigma}+\mu)G_{ff,k\sigma} &= 1-c\chi_0 G_{cf,k\sigma}, \\
(z-\epsilon_{k\sigma}+\mu)G_{cf,k\sigma} &= -c\chi_0 G_{ff,k\sigma}, \\
(z-\lambda_{\sigma}+\mu)G_{fc,k\sigma} &= -c\chi_0 G_{cc,k\sigma},
\end{split}
}
and consequently
\beq{
(z-\lambda_{\sigma}+\mu)^2 G_{ff,k\sigma} = (c\chi_0)^2 G_{cc,k\sigma}.
}
In this approach, writing $z=\omega+i\delta$, the Fermi level
corresponds to $\omega=0$. We use a different convention. We absorb
$\mu$ into $z$: $\tilde z=z+\mu$.  Also the Green's functions take $\tilde z$ as their
argument. With this choice, spectral functions are obtained with
replacement $\tilde z=\omega+i\delta$ and there are no explicit $\mu$
in the expressions for Green's functions. $\mu$ only appears as an integration limit
(or in the Fermi-Dirac distribution). We drop writing the tilde in $\tilde z$ in the following.

The quasiparticle band edges are
\beq{
\begin{split}
\omega_{1,\sigma} &= \frac{1}{2}
\left(
\epsilon_\sigma+\lambda_\sigma-D-\sqrt{(\epsilon_\sigma-\lambda_\sigma-D)^2+4c^2\chi_0^2}
\right),\\
\omega_{2,\sigma} &= \frac{1}{2}
\left(
\epsilon_\sigma+\lambda_\sigma+D-\sqrt{(\epsilon_\sigma-\lambda_\sigma+D)^2+4c^2\chi_0^2}
\right),\\
\omega_{3,\sigma} &= \frac{1}{2}
\left(
\epsilon_\sigma+\lambda_\sigma-D+\sqrt{(\epsilon_\sigma-\lambda_\sigma-D)^2+4c^2\chi_0^2}
\right),\\
\omega_{4,\sigma} &= \frac{1}{2}
\left(
\epsilon_\sigma+\lambda_\sigma+D+\sqrt{(\epsilon_\sigma-\lambda_\sigma+D)^2+4c^2\chi_0^2}
\right).
\end{split}
}
In the multiindex $(i,\sigma)$, $\sigma$ is spin, while $i$ enumerates the
band edges from the lowest to the highest. Furthermore 
\beq{
\epsilon_{\sigma}=J \tilde m_f \frac{\sigma}{2} + \mu_B g_c H
\frac{\sigma}{2} = \mu_B g_c \tilde H_c \frac{\sigma}{2}.
}

The final closed-form expressions for the spectral functions are
\beq{
\rho_{c,\sigma}(\omega) = \rho_c^0 \sum_{i=1}^4 (-1)^{i-1}
\theta(\omega-\omega_{i,\sigma}),
}
\beq{
\label{Sprhof}
\rho_{f,\sigma}(\omega) = \frac{(c \chi_0)^2}{(\omega-\lambda_\sigma)^2}
\rho_{c,\sigma}(\omega).
}
We also have
\beq{
\rho_{cf,\sigma}(\omega) = -\frac{c\chi_0}{\omega-\lambda_\sigma}
\rho_{c,\sigma}(\omega).
}

The energy eigenvalues are
\beq{
E_{k,\sigma} = \frac{1}{2} \left(
\epsilon_{k,\sigma} + \lambda_\sigma \pm
\sqrt{(\epsilon_{k,\sigma}-\lambda_\sigma)^2 + 4c^2\chi_0^2}
\right).
}

\subsubsection{Mean-field equations}

We can derive the system of mean-field equation using the fluctuation-dissipation theorem at $T=0$:
\beq{
\begin{split}
\ave{A B} 
&= -\int_{\mu}^{\infty} \frac{d\omega}{\pi} G''_{AB}(\omega) \\
&= \int_{-\infty}^\mu d\omega \rho_{BA}(\omega).
\end{split}
} 
We obtain
\begin{align}
& n_c=\sum_{\sigma} \ave{c^{\dagger}_{\sigma} c_{\sigma}}=\sum_{\sigma} \int d\omega \rho_{c,\sigma}(\omega), \\
& 1=n_f=\sum_{\sigma} \ave{f^{\dagger}_{\sigma} f_{\sigma}}=\sum_{\sigma} \int d\omega \rho_{f,\sigma}(\omega), \\
& \tilde m_c=1/2 \sum_{\sigma} \sigma \ave{c^{\dagger}_{\sigma}
c_{\sigma}}=1/2 \sum_{\sigma} \int \sigma d\omega \rho_{c,\sigma}(\omega), \\
& \tilde m_f=1/2 \sum_{\sigma} \sigma \ave{f^{\dagger}_{\sigma}
f_{\sigma}}=1/2 \sum_{\sigma} \int \sigma d\omega \rho_{f,\sigma}(\omega).
\label{Eq.:system05}
\end{align}
In all integrals, the lower integration limit is $-\infty$, while the
upper is the chemical potential $\mu$.

For the gap equation we take the symmetrized spectral function
$$A_{cf,\sigma}=-\frac{1}{2 \pi}[\mathrm{Im}
G_{cf}(\omega+i\delta)+\mathrm{Im}
G_{fc}(\omega+i\delta)]=\rho_{cf,\sigma}.$$ 
This gives:
\begin{align}
&\frac{1}{2} \ave{f_{\uparrow}^{\dagger}c_{\uparrow}+c_{\uparrow}^{\dagger}f_{\uparrow}+f_{\downarrow}^{\dagger}c_{\downarrow}+c_{\downarrow}^{\dagger} f_{\downarrow}}=
\frac{1}{2}2 \sqrt{2}\ave{\chi_{0}}= \\
&\sum_{\sigma} \int d\omega A_{fc,\sigma}(\omega)=-c \chi_{0}\sum \int
d\omega \frac{1}{\omega-\lambda_\sigma} \rho_{c,\sigma}(\omega)
\end{align}

We now assume $\chi_0 \neq 0$. Using $c=3J/(4\sqrt{2})$, we finally
find the gap equation
\beq{
\boxed{
\sum_{\sigma}\int_{-\infty}^{\mu} d\omega
\frac{\rho_{c,\sigma}(\omega)}{\omega-\lambda_{\sigma}}=-8/3J.}
\label{Eq.:gap05}
}
This set of non-linear equations had been previously derived in
Refs.~\cite{beach2005,kusminskiy2008}, while in Ref. \cite{li2010} a
somewhat different mean-field decoupling was used.

\subsubsection{Evaluation of energy}

\newcommand{\bM}{\mathbf{M}}
\newcommand{\bA}{\mathbf{A}}
\newcommand{\bG}{\mathbf{G}}

The total energy can be evaluated as
\beq{
\begin{split}
E_\mathrm{GS}
&= \left\langle H_\mathrm{MF} \right\rangle \\
&= \sum_k \left\langle \sum_{ij} c_{ki}^\dag c_{kj} M_{k,ij} + E_0
\right\rangle \\
&= \sum_k \left( \sum_{ij} \int_{-\infty}^\mu
A_{\epsilon_k,ij}(\omega) M_{\epsilon_k,ij} \mathrm{d}\omega + E_0 \right).
\end{split}
}
We used a symmetrized spectral function
\beq{
A_{ij}(\omega) = \frac{1}{2} \left[
-\frac{1}{\pi} \mathrm{Im} G_{ij}(\omega+i \delta)
-\frac{1}{\pi} \mathrm{Im} G_{ji}(\omega+i \delta)
\right],
}
since
\beq{
\int_{-\infty}^\mu A_{ij}(\omega)\mathrm{d}\omega = \frac{1}{2} \expv{ c^\dag_{i} c_j + c^\dag_{j} c_i }.
}
Then
\beq{
\begin{split}
\frac{E_\mathrm{GS}}{N} 
&=
E_0 + \int_{-D}^{D} \rho(\epsilon)d\epsilon \int_{-\infty}^\mu
\mathrm{Tr}[\mathbf{A}_{\epsilon}(\omega) \mathbf{M}_{\epsilon}]
\mathrm{d}\omega.
\end{split}
}
Note that both $\mathbf{A}$ and $\mathbf{M}$ have out-of-diagonal
matrix elements.
Now we use
\beq{
\Tr[\bA(\omega)\bM]
=-\frac{1}{\pi} \mathrm{Im}\Tr[\bG(\omega+i\delta)\bM]
=-\frac{1}{\pi} \mathrm{Im}\Tr[(\omega+i\delta-\bM)^{-1}\bM]
=-\frac{1}{\pi} \mathrm{Im}\Tr[(\omega+i\delta-\bM)^{-1}\omega]
=\Tr[\bA(\omega)]\omega,
}
which follows from the fact that $\mathrm{Im}[1/(z-x)]$ is a delta
distribution, and we have used a transformation to the eigenbasis
and back to replace $\bM$ by $\omega$ in the third step.
Thus, after the integration over $\epsilon$,
\beq{
\label{eq33}
\frac{E_\mathrm{GS}}{N} 
= E_0 + \sum_\sigma \int_{-\infty}^\mu \omega d\omega 
\left[ \rho_{c,\sigma}(\omega) + \rho_{f,\sigma}(\omega) \right].
}

We also have
\beq{
\begin{split}
\frac{N_c+N_f}{N} &= \int_{-D}^D \rho(\epsilon)d\epsilon
\int_{-\infty}^\mu \Tr[A_{\epsilon}(\omega)] d\omega \\
&= \sum_\sigma \int_{-\infty}^{\mu} \left[ \rho_{c,\sigma}(\omega) +
\rho_{f,\sigma}(\omega) \right] d\omega,
\end{split}
}
thus finally,
\beq{
\frac{K_\mathrm{GS}}{N} = E_0 +
\sum_\sigma \int_{-\infty}^\mu d\omega 
(\omega-\mu) \left[
\rho_{c,\sigma}(\omega) + \rho_{f,\sigma}(\omega) \right].
\label{Eq:ener}
}

\begin{figure}
\includegraphics[width=0.6\textwidth]{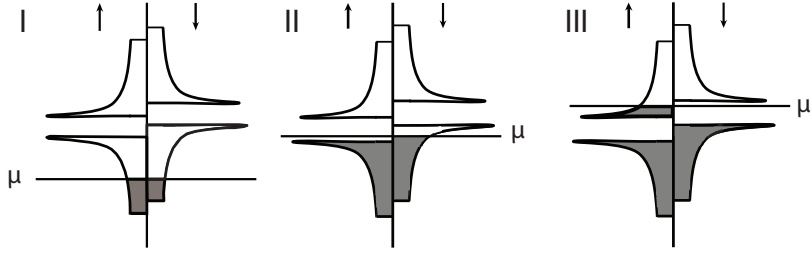}
\caption{Sketch of the possible placements of the bands with respect to the chemical potential. The phase I correspond to the phase B',
 phase II to the phase A and phase III to the phase B in the DMFT calculations.}
\label{Fig.:skica}
\end{figure}

We would like to evaluate Eq.~\eqref{Eq:ener} for two different cases
represented on Fig.~\ref{Fig.:skica}, namely cases I and II:
\beq{
\begin{split}
\frac{K_{GS}}{N}&=E_0 +\sum\limits_{\sigma} \int_{-\infty}^{\mu} d\omega (\omega-\mu) 
\left( 1+\frac{(c\chi_{0})^2}{(\omega-\lambda_{\sigma})^2} \right) \rho_{c\sigma}(\omega) \\
&=
E_0+E_c+(c\chi_0)^2\sum_\sigma \int_{-\infty}^{\mu} d\omega \frac{\omega-\lambda_{\sigma}+\lambda_{\sigma}-\mu}{(\omega-\lambda_{\sigma})^2} \rho_{c\sigma}(\omega)\\
&=
E_0+E_c+(c\chi_0)^2 \sum\limits_{\sigma} \int_{-\infty}^{\mu}d\omega \frac{\rho_{c,\sigma}(\omega)}{\omega-\lambda_{\sigma}}+\sum\limits_{\sigma}(\lambda_{\sigma}-\mu)n_{f,\sigma}\\
&=
E_0+E_c+(c\chi_0)^2(-8/3J)+\sum\limits_{\sigma}(\lambda_{\sigma}-\mu)n_{f,\sigma},
\end{split}
}
where $E_c=\int_{-\infty}^{\mu}d\omega (\omega-\mu)
\rho_{c,\sigma}(\omega)$ and in the last line we have use the gap
equation, see Eq.~\eqref{Eq.:gap05}. We need to evaluate
\beq{
\begin{split}2\sum_\sigma
(\lambda_{\sigma}-\mu) n_{f,\sigma} &= \sum_\sigma \left( \lambda-\mu +
\mu_B g_f \tilde H_f \frac{\sigma}{2} \right) n_{f\sigma} \\
&= (\lambda-\mu) n_f + 2\mu_B g_f \tilde H_f \tilde m_f.
\end{split}
}

For $H=0$, this is equal to

\beq{
(\lambda-\mu) n_f +  J \tilde m_c \tilde m_f.
\label{Eq.:nfrel}
}

Case I is when $\omega_{1,\sigma}<\mu<\omega_{2,\sigma}$ for both spin
orientations. We can write:
\beq{
E_c=\rho_{c,0}\sum\limits_{\sigma}[(\mu^2-\omega_{1,\sigma}^2)/2]-\mu
n_c
}
and
\beq{
\begin{split}
\frac{K_{GS}}{N}&=\frac{3}{4} J \chi_0^2-J \tilde m_c \tilde m_f-\lambda+ \rho_{c,0}\sum\limits_{\sigma}[(\mu^2-\omega_{1,\sigma}^2)/2]-\mu n_c-\frac{3}{4} J \chi_0^2+\sum\limits_{\sigma}(\lambda_{\sigma}-\mu)n_{f,\sigma}\\
&=
-J \tilde m_c m_f-\lambda+ \rho_{c,0}\sum\limits_{\sigma}[(\mu^2-\omega_{1,\sigma}^2)/2]-\mu n_c + (\lambda-\mu)n_f + J \tilde m_c \tilde m_f \\
&=
 \rho_{c,0}\sum\limits_{\sigma}[(\mu^2-\omega_{1,\sigma}^2)/2]-\mu [n_c+n_f]
\end{split}
}
where in the second line  we have used  Eq.~\eqref{Eq.:nfrel} and in the last line $\ave{n_f}=1.$

Case II is when $\omega_{2,\uparrow}<\mu<\omega_{3,\uparrow}$ and only difference is 
that $\mu\rightarrow \omega_{2,\uparrow}$ in integration limit for $\uparrow$ c electrons.
Therefore, the only difference is in the evaluation of  $E_c$:
\beq{
E_c=
\rho_{c}^0 
\left(
\frac{\mu^2-\omega_{1,\downarrow}^2}{2}
+
\frac{\omega_{2,\uparrow}^2-\omega_{1,\uparrow}^2}{2}
\right) - \mu n_c.
\label{Eq.:kfree1}
}

%

\subsection{The $S=1$ case}

We next proceed with an analogous treatment for the $S=1$ problem. We
decompose the interaction term into doublet and quadruplet terms,
$1/2\otimes 1=1/2 \oplus 3/2$. We find: 
\begin{equation}
\label{eq79}
\vc{s} \cdot \vc{S} =
-\sum\limits_{i=1}^{2} \chi_{d,i}^{\dagger}\chi_{d,i}+(1/2)\sum\limits_{i=1}^{4}\chi_{q,i}^{\dagger}\chi_{q,i},
\end{equation}
where $\chi_{d,i}(\chi_{q,j})$ are the doublet ($i=1,2$) and
the quadruplet ($j=1,2,3,4$) sets of operators under the spin
$SU(2)$ symmetry, namely:
\beq{
\chi_{d,1}=-\sqrt{1/3}c_{\downarrow}^{\dagger}f_{0}-\sqrt{2/3}c_{\uparrow}^{\dagger}
f_{1},  \qquad \chi_{d,2}=
\sqrt{2/3}c_{\downarrow}^{\dagger}f_{-1}+\sqrt{1/3}c_{\uparrow}^{\dagger}
f_{0}, }
\beq{
\chi_{q,1}= - c_{\uparrow}^{\dagger} f_{-1}, \quad
\chi_{q,2}=-\sqrt{1/3}c_{\downarrow}^{\dagger}f_{-1}+\sqrt{2/3}c_{\uparrow}^{\dagger}f_{0},
\quad
\chi_{q,3}=\sqrt{2/3}c_{\downarrow}^{\dagger}
f_0-\sqrt{1/3}c_{\uparrow}^{\dagger} f_1, \quad 
\chi_{q,4}=- c_{\downarrow}^{\dagger} f_{1}.
}
These operators again form a complete set in the spin sector. The
decomposition in Eq.~\eqref{eq79} is exact.

We focus on the doublet part and set all quadruplet fields to zero,
$\expv{\chi_{q,j}}=0.$
We explicitly break the
SU(2) symmetry by setting $\expv{\chi_{d,2}}=0$ and use the $U(1)$ gauge
freedom to make $\expv{\chi_{d,1}}$ real. In analogy with the $S=1/2$ case,
we make a second mean-field decomposition in the magnetic channel.
The mean-field Hamiltonian has a simple wave-vector representation:

\begin{equation}
\mathcal K_{MF}=\sum\limits_{k}
\left(
c_{k,\downarrow}^{\dagger}\, 
c_{k,\uparrow}^{\dagger}\,
f_{k,-1}^{\dagger}\,
f_{k,0}^{\dagger}\,
f_{k,+1}^{\dagger}
\right) \tilde M_{k}
\begin{pmatrix}
 c_{k,\downarrow} \\
 c_{k,\uparrow} \\
 f_{k,-1} \\
 f_{k,0} \\
 f_{k,+1} 
  \end{pmatrix} + \sum_k E_0,
  \label{Eq.:ham}
\end{equation}
where the matrix $\tilde M_{k}$ is
%
%
\begin{equation}
\begin{pmatrix}
 \epsilon_{k,\downarrow} -\mu & 0 & 0 & J\chi_{d,1}/\sqrt{3} & 0 \\
 0 & \epsilon_{k,\uparrow}-\mu & 0 & 0  & \sqrt{2/3} J \chi_{d,1} \\
 0 & 0 & \lambda_{-1}-\mu & 0 & 0 \\
 J \chi_{d,1}/\sqrt{3} & 0 & 0 & \lambda_{0}-\mu & 0 \\
 0 & \sqrt{2/3} J \chi_{d,1} & 0 & 0 & \lambda_{1}-\mu
  \end{pmatrix}
\end{equation}
and 

\beq{
\epsilon_{k,\sigma}=\epsilon_{k} + J \tilde m_f \frac{\sigma}{2} + \mu_B g_c H
\frac{\sigma}{2} =
\epsilon_{k}+ \mu_B g_c \tilde H_c \frac{\sigma}{2},
}
\beq{
\lambda_{i}
= \lambda + J \tilde m_c i + \mu_B g_f H i =
\lambda+ \mu_B g_f \tilde H_f i,
}
\beq{
E_0=J\chi_{d,1}^2- J \tilde m_c \tilde m_f-\lambda,
}
with $\sigma=\pm 1$, $i=-1,0,1$. The effective field felt by the $c(f)$
electrons is given by 
\beq{
\tilde H_{c(f)}=H+\frac{J \tilde
m_{f(c)}}{\mu_B g_{c(f)}}.
}
%

 
 The EOMs are
  \beq{
 \begin{split}
 (z-\epsilon_{k,\downarrow}+\mu)&G_{c_{\downarrow} ,k} = 1+\sqrt{\frac{1}{3}} J \chi_{d1} G_{f_0,c_{\downarrow},k} \\
 (z-\epsilon_{k,\uparrow}+\mu)&G_{c_{\uparrow} ,k} = 1+\sqrt{\frac{2}{3}} J \chi_{d1} G_{f_1,c_\uparrow,k} \\
 (z-\lambda_{-1}+\mu)&G_{f_{-1} ,k} = 1 \\
  (z-\lambda_0+\mu)&G_{f_0,c_{\downarrow} ,k} = \sqrt{\frac{1}{3}} J \chi_{d1} G_{c_\downarrow,c_{\downarrow},k} \\
   (z-\epsilon_{k,\downarrow}+\mu)&G_{c_{\downarrow}f_{0} ,k} = \sqrt{\frac{1}{3}} J \chi_{d1} G_{f_0,f_0,k} \\
 \end{split}
 }
  and note also that $G_{ij,k}(z)=G_{ji,k}(z),$ while for the diagonal elements we used $G_{ii,k}(z)=G_{i,k}(z).$ Consequently
  \beq{
  \begin{split}
  (z-\lambda_{0}+\mu)^2 G_{f_0  ,k} & =(z-\lambda_0 +\mu)\left( 1+
  \frac{(J \chi_{d1})^2}{3} G_{c_\downarrow ,k} \right)\\
  (z-\lambda_{1}+\mu)^2 G_{f_1  ,k} & =(z-\lambda_1 +\mu)\left( 1+
  \frac{2(J \chi_{d1})^2}{3} G_{c_\uparrow ,k} \right).
  \end{split}
  }
 Once more we absorb $\mu$ into $z$: $\tilde z= z +\mu $ and drop writing tilde in $\tilde z$ in the following.
 The quasiparticles band edges $\omega_{i,\sigma}$ are:
 \begin{align}
   &\omega_{1,\sigma}= \left( 3 \epsilon_{\sigma}-3D+3\lambda_{\sigma}-
 \sqrt{9(\epsilon_{\sigma}-D-\lambda_{\sigma})^2+12 F_\sigma (J \chi_{d,1})^2} \right) /6 \nonumber \\
  &\omega_{2,\sigma}=\left(3 \epsilon_{\sigma}+3D+3\lambda_{\sigma}-
 \sqrt{9(\epsilon_{\sigma}+D-\lambda_{\sigma})^2+12 F_\sigma (J \chi_{d,1})^2}\right)/6 \nonumber \\
  &\omega_{3,\sigma}=\left(3 \epsilon_{\sigma}-3D+3\lambda_{\sigma}+
  \sqrt{9(\epsilon_{\sigma}-D-\lambda_{\sigma})^2+12 F_\sigma (J \chi_{d,1})^2}\right)/6 \nonumber \\
   &\omega_{4,\sigma}=\left(3 \epsilon_{\sigma}+3D+3\lambda_{\sigma}+
   \sqrt{9(\epsilon_{\sigma}+D-\lambda_{\sigma})^2+12 F_\sigma (J \chi_{d,1})^2}\right)/6,
   \end{align}
 where $\epsilon_\sigma$ has been defined in the section on the $S=1/2$ model, while
 \beq{
 F(1)=2, \quad F(-1)=1.
 }
 and, furthermore,
 \beq{
 \lambda_{\downarrow} = \lambda_0, 
 }
 \beq{
 \lambda_{\uparrow}=\lambda_1.
 }
 
The spectral functions are given by:
 \beq{
  \rho_{c, \sigma}(\omega)=\rho_{c}^{0} \sum\limits_{i=1}^{4} (-1)^{i-1}
  \theta(\omega-\omega_{i,\sigma}),
 }
 \beq{
  \rho_{f, -1}(\omega)=\delta(\omega-\lambda_{-1}),
  }
  \beq{
 \rho_{f,0}(\omega)=F_{-1} \frac{(J \chi_{d,1})^2}{3(z-\lambda_{0})^2} \rho_{c,\downarrow}(\omega),
 }
 \beq{
 \rho_{f,1}(\omega)=F_{1} \frac{(J \chi_{d,1})^2}{3(z-\lambda_{1})^2} \rho_{c,\uparrow}(\omega).
 }
\beq{
 \rho_{f_1c_\uparrow}(\omega)=\sqrt{\frac{2}{3}} \frac{(J \chi_{d,1})}{(\omega-\lambda_{1})} \rho_{c_\uparrow}(\omega).
 }
\beq{
 \rho_{f_0c_\downarrow}(\omega)=\sqrt{\frac{1}{3}} \frac{(J \chi_{d,1})}{(\omega-\lambda_{0})} \rho_{c_\downarrow}(\omega).
 }
The $f_{-1}$ must be unoccupied, otherwise the number of $f$ electrons
cannot be exactly 1. Thus $\lambda_{-1}>\mu$. This also implies
that $f_{-1}$ must be the highest in energy of the $f$ states,
thus $\tilde m_c<0$ and consequently $\tilde m_f>0$.

\subsubsection{The mean-field equations}

Using the fluctuaction-dissipation theorem at $T=0$,
we find
\begin{align}
& n_c=\sum_{\sigma} \ave{c^{\dagger}_{\sigma} c_{\sigma}}=\sum_{\sigma} \int d\omega \rho_{c,\sigma}(\omega), \\
& 1=n_f=\sum_{i} \ave{f^{\dagger}_{i} f_{i}}=\sum_{i} \int d\omega \rho_{f,i}(\omega), \\
& \tilde m_c=1/2 \sum_{\sigma} \sigma \ave{c^{\dagger}_{\sigma}
c_{\sigma}}=1/2 \sum_{\sigma} \int \sigma d\omega \rho_{c,\sigma}(\omega), \\
& \tilde m_f= \sum_{i} i \ave{f^{\dagger}_{i} f_{i}}=\sum_{i} i \int d\omega \rho_{f,i}(\omega).
\label{Eq.:system1}
\end{align}

For the gap equation we take symmetrized spectral function 
$A_{c\sigma; f i}=-\frac{1}{2 \pi}[\mathrm{Im}
G_{c\sigma;fi}(\omega+i\delta)+\mathrm{Im} G_{fi;c\sigma}(\omega+i\delta)]=\rho_{c\sigma;f,i},$ where 
$G_{c\sigma;fi}(z)=\langle\langle c^{\dagger}_{\sigma}; f_{i}
\rangle\rangle_z$, etc. For the evaluation of $\ave{\chi_1}$ we will
need two off-diagonal spectral functions:
\begin{align}
 \rho_{c\downarrow;f0}=\frac{J \chi_1}{\sqrt{3}(z-\lambda_0)}\rho_{c\downarrow}(\omega) \\
 \rho_{c\uparrow;f1}=\frac{\sqrt{2/3} J \chi_1}{z-\lambda_1}\rho_{c\uparrow}(\omega).
\end{align}
The expectation value is
\begin{align}
&\ave{\chi_1}=\frac{1}{2}[ -\sqrt{1/3} (\ave{c_{\downarrow}^{\dagger}f_{0}+f_{0}^{\dagger}c_{\downarrow}})
-\sqrt{2/3}( c_{\uparrow}^{\dagger}f_{1}+f_{1}^{\dagger} c_{\uparrow})]= 
-\sqrt{1/3} \int A_{c\downarrow,f0}(\omega)d\omega-\sqrt{2/3} \int A_{c\uparrow,f1}(\omega) d\omega \\
&= -\sqrt{1/3} \frac{J \chi_1}{\sqrt{3}} \int d\omega \rho_{c\downarrow}/(\omega-\lambda_0)-\sqrt{2/3}\sqrt{2/3}J \chi_1 \int d\omega
\rho_{c,\uparrow}(\omega)/(\omega-\lambda_1) \\
&=-(J \chi_1/3) \left[
\int d\omega \rho_{c,\downarrow}(\omega)/(\omega-\lambda_0)
\right]-2J\chi_1/3
\left[
\int d\omega \rho_{c\uparrow}(\omega)/(\omega-\lambda_1)
\right].
\end{align}
Finally, we obtain the gap equation:
\begin{equation}
\boxed{\int d\omega
\rho_{c\downarrow}(\omega)/(\omega-\lambda_{0})+2\int d\omega\rho_{c\uparrow}(\omega)/(\omega-\lambda_1)=-3/J.}
\label{Eq.:gap1}
\end{equation}
This equation has essentially the same structure as the gap equation
for the $S=1/2$ case.

\subsubsection{Evaluation of energy}

The total energy can be evaluated in analogy to the $S=1/2$ case. We
find
\beq{
\frac{{K_\mathrm{GS}}}{N}=E_{0}+\sum\limits_{\sigma} \int_{-\infty}^{\mu} d\omega  (\omega-\mu) [\rho_{c,\sigma}(\omega)+\rho_{f,i(\sigma)}(\omega)]+(\lambda_{-1}-\mu) \theta(\omega-\lambda_{-1})
\label{Eq.:ene1}
}
We evaluate Eq.~\ref{Eq.:ene1} for two different cases represented on
Fig.~\ref{Fig.:skica}. For case I, 
$E_c=\rho_{c,0}\sum\limits_{\sigma}[(\mu^2-\omega_{1,\sigma}^2)/2]-\mu
n_c$, thus
\beq{
\begin{split}
\frac{K_{GS}}{N}&=\rho_{c,0}\sum\limits_{\sigma}[(\mu^2-\omega_{1,\sigma}^2)/2]+\lambda (n_{f0}+n_{f1})-\mu (n_f+n_c)
\end{split}
}

In the case II,
\beq{
E_c=\rho_{c}^0 
\left( 
\frac{\mu^2-\omega_{1,\downarrow}^2}{2}
+
\frac{\omega_{2,\uparrow}^2-\omega_{1,\uparrow}^2}{2}
\right) - \mu n_c.
}

\section{Phase diagrams for $S=1/2$ and $S=1$}

We now discuss the different possible mean-field phases for the
$S=1/2$ and $S=1$ Kondo lattice models. 

One possible phase is a pure saturated ferromagnetic phase with
magnetization $m_f=-g_f \mu_B S$ and with zero hybridisation,
$\chi_{d,1}=0$.  If conducting electrons are completely polarized we
call it the polar phase  and the magnetization of conducting
electrons is then given by
\beq{
m_{c,P-I}=\mu_B g_c n_c/2.
}
%
%
%

%
%
%

\begin{figure}
  \includegraphics[width=0.35\textwidth]{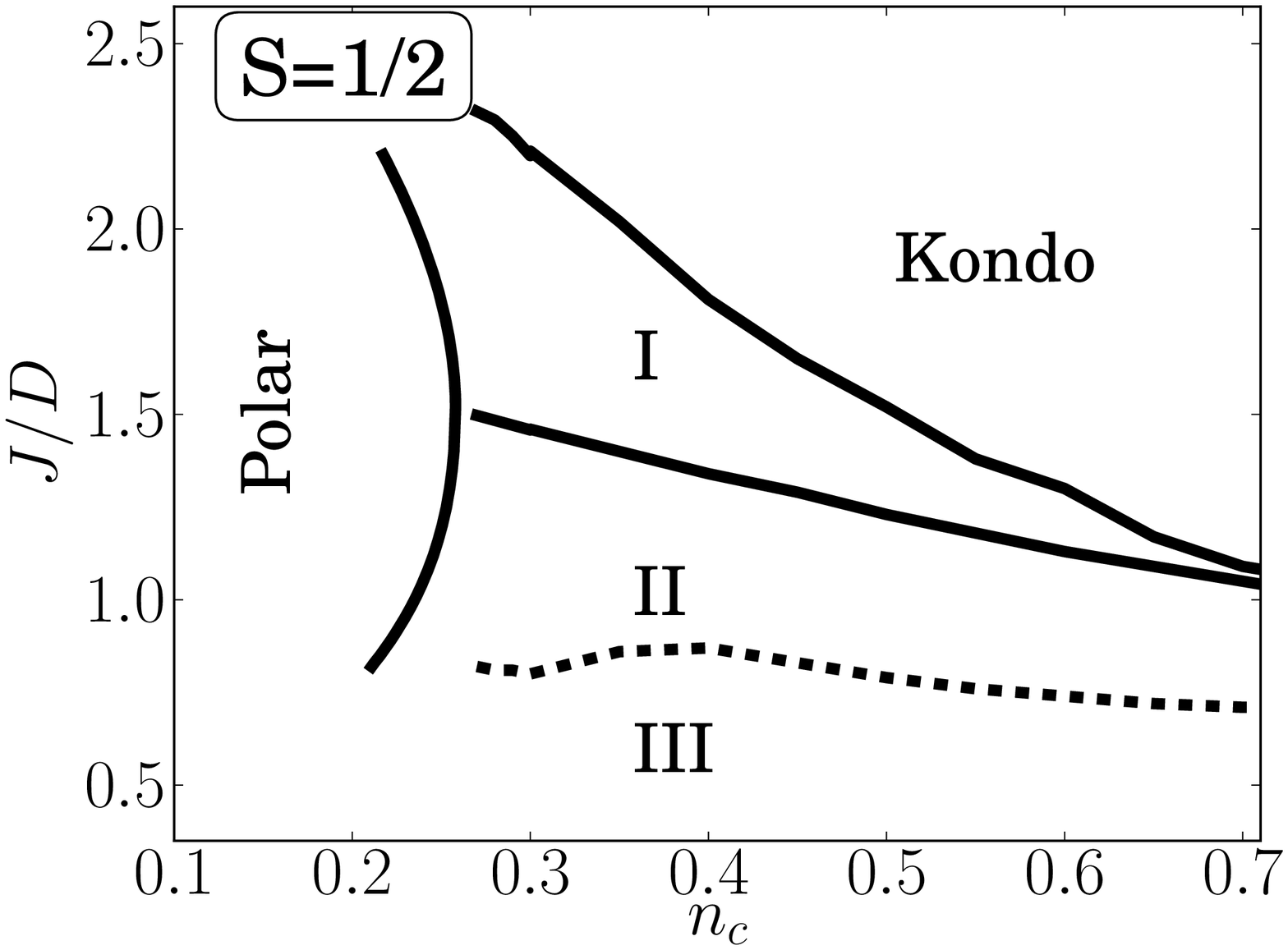}
  \includegraphics[width=0.35\textwidth]{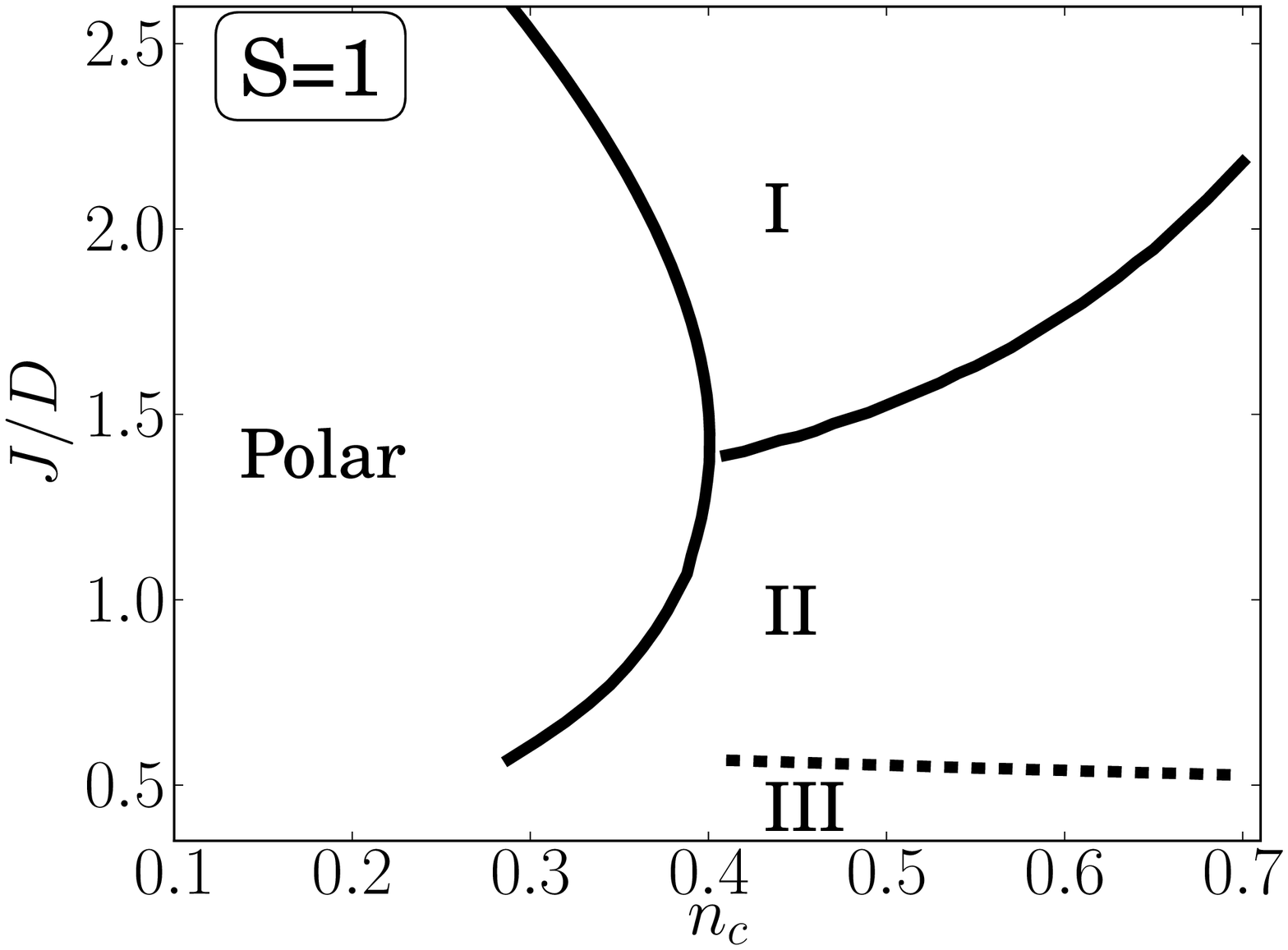} 
  \caption{Ground state phase diagram of the KLM: (a) $S=1/2$, (b)
  $S=1$ with Land\'{e} factors $g_c=g_f=1.$ For the description of
  phases I, II, see the discussion in the text and Fig.~\ref{Fig.:skica}.
   Phase Polar-I represent polarized phase with zero hybridisation and Kondo phase is paramagnetic phase ($m_c=m_f=0.$). The dashed line represent the transition into phase where we could not find the convergent solution,
   but phase III is expected, see discussion in the text.
   }
  \label{Fig:diagram}
\end{figure}

For intermediate coupling regime, we distinguish between the
ferromagnetic phases I, II, and III, which all have a finite value of
the hybridisation parameter $\chi_{d,1}$ (thus a spectral gap). They
are schematically represented in in Fig.~\ref{Fig.:skica}. The phase I
with the electron pockets, corresponds to the phase B' in the DMFT
calculations. The phase II with the chemical potential in the gap
corresponds to the DMFT phase A.  The numerical results in the phase
II clearly indicate that as we lower $J$ the transition into phase III
is expected, but when $\mu>\omega_{3,\uparrow}$ we were not able to
find convergent solution in the regime of small $J$, as marked by the
the dashed line in Fig.~\ref{Fig:diagram}, see also
\cite{kusminskiy2008}. This phase III would correspond to the phase B
in the DMFT calculations, where this is a stable phase.
  
The phase boundary between the phase I and II or between the phase II
and III is given by the condition
\beq{
\tilde m_c+\tilde m_f=(2S-n)/2,
}
for the expectation values of spin $z$ component, which shows plateau
behaviour irrespective of the Land\'{e} factors or, equivalently,
\beq{m_c/g_c+m_f/g_f=-\mu_B (2S-n)/2.}
This is equivalent to the condition that
\beq{
\mu=\omega_{2(3),\uparrow}
}
for transition between the phases I$\rightarrow$ II (II$\rightarrow$ III).

The pure Kondo singlet (paramagnetic) phase is defined by
$m_c=0,m_f=0,\chi_{d,1}\neq 0$. We only find it for $S=1/2$. In the
$S=1$ model the hole pocket never emerges; instead, the chemical
potential becomes attached near the top of the bottom band for large
$J$. In fact, similar behavior is also observed in the DMFT solutions.
The boundary between phase I and the Kondo phase
is determined by the condition 
\beq{
m_{f}=m_c=0.
}

The boundary between the phases I,II and polar-I is given by the condition 
\beq{
\chi_{d,1}=0.
}

We conclude that the qualitative features of the static MF and DMFT
phase diagrams are rather similar, except that in the static
mean-field theory the phase III is not stable. The main difference
compared to previous works \cite{Beach:2004bf,kusminskiy2008,li2010}
is the finding that in the MF treatment the metamagnetic transition is
described by the transition I $\rightarrow$ II, while in the DMFT
there are two different scenarios for metamagnetic transitions, either
the transition I $\rightarrow$ II or the transition II $\rightarrow$
III, where only the former is expected for physically relevant model
parameters.

\end{document}